\documentclass{article}%
\usepackage{amsmath}
\usepackage{amsfonts}
\usepackage{amssymb}
\usepackage{graphicx}%
\providecommand{\U}[1]{\protect\rule{.1in}{.1in}}
\begin{document}

\title{Entropic Dynamics approach to Relational Quantum Mechanics}
\author{Ariel Caticha and Hassaan Saleem\\{\small Department of Physics, University at Albany--SUNY, Albany, NY 12222,
USA}}
\date{}
\maketitle

\begin{abstract}
The general framework of Entropic Dynamics (ED) is used to construct
non-relativistic models of relational quantum mechanics from well known
inference principles --- probability, entropy and information geometry.
Although only partially relational --- the absolute structures of simultaneity
and Euclidean geometry are still retained --- these models provide a useful
testing ground for ideas that will prove useful in the context of more
realistic relativistic theories. The fact that in ED the positions of
particles have definite values, just as in classical mechanics, has allowed us
to adapt to the quantum case some intuitions from Barbour and Bertotti's
classical framework. Here, however, we propose a new measure of the mismatch
between successive states that is adapted to the information metric and the
symplectic structures of the quantum phase space. We make explicit that ED is
temporally relational and we construct non-relativistic quantum models that
are spatially relational with respect to rigid translations and rotations. The
ED approach settles the longstanding question of what form should the
constraints of a classical theory take after quantization: the quantum
constraints that express relationality are to be imposed on expectation
values. To highlight the potential impact of these developments, the
non-relativistic quantum model is parametrized into a generally covariant form
and we show that the ED approach evades the analogue of what in quantum
gravity has been called the problem of time.

\end{abstract}

\section{Introduction}

Many, perhaps all, the difficulties in the study of space and of time can be
traced to their invisibility. We do not see space; we see matter in space. We
do not see time; we see changes in things. Although these studies have
resulted in an increased understanding of the properties of matter in
space-time the problem remains that it is not always clear how to disentangle
which properties should be attributed to matter and which to space. Newton
managed to evade the issue and created a successful science of mechanics but
only at the cost of adopting an absolute concept of motion relative to an
abstract \textquotedblleft mathematical\textquotedblright\ space and time.
Leibniz objected that such absolute positions would not be observable and
therefore not real. Instead, he advocated for a concept of motion defined in
terms of the observable relative distances. Newton had, however, a definite
dynamical theory to back him up, while Leibniz had no such support. As a
result Newton won the day, and the relational argument lay largely dormant for
almost two centuries until revived by Mach and specially by Einstein. In the
latter decades of the 20th century, as a stepping stone towards a quantum
theory of gravity, the thrust towards a fully relational dynamics was once
again resumed in the form of Hamiltonian formulations of general relativity
\cite{Dirac 1958}-\cite{Hojman Kuchar Teitelboim 1976}. More recently the
understanding of relational motion achieved a certain level of completion ---
at least within the context of classical physics --- with the explicitly
Machian line of research that originates with J. Barbour and B. Bertotti
\cite{Barbour Bertotti 1982}-\cite{Barbour 2010}.

Barbour and Bertotti's basic insight \cite{Barbour Bertotti 1982} was that a
formulation in terms of the relative interparticle distances is not practical.
Instead one should focus attention on the fact that two configurations that
differ by arbitrary rigid (\emph{i.e.}, global) displacements and/or rigid
rotations describe exactly the same physical situation. Thus, in formulating a
relational dynamics, whatever measure one adopts to quantify the change from
one configuration to the next, the actual intrinsic change should remain
unaffected by independent rigid shifts and rotations of each configuration.
This is achieved through a technique Barbour and Bertotti called
\textquotedblleft best matching\textquotedblright\ (BM). The idea is to
introduce a quantitative measure of the \textquotedblleft
mismatch\textquotedblright\ between two successive configurations and then
shift and rotate one configuration \emph{relative} to the other to find the
location that minimizes the mismatch. Thus, motion is not defined as relative
to an absolute space, but relative to the earlier state of the system itself.
Barbour and Bertotti's intrinsic change and best matching have added an
interesting twist: what is relational is not the notion of space but the
notion of change.

With Quantum Mechanics (QM), however, new problems arise. In its standard
Copenhagen interpretation QM is manifestly non-relational. It lives in
Newton's absolute space and time, or at best, in Minkowski's absolute
space-time, and its particles do not have definite positions much less
definite relative distances. But then the interpretations of QM have always
been a source of controversy and many have been proposed. One approach
inspired by Barbour and Bertotti is due to Gryb \cite{Gryb 2009}. Another
relevant example, Rovelli's relational interpretation, is tailored to
formulating a background-independent quantum theory of gravity \cite{Rovelli
1996}{}\cite{Rovelli 2009}. We shall follow a different path. First, however,
we must clarify a potential of confusion between this line of research and the
different line of research by the same name pursued by Rovelli and
collaborators (see \cite{Rovelli 1996}\cite{Rovelli 2021}). As we pointed out
above we use the term `relational' in a sense that can be traced historically
in the 18th century back to Newton and Leibniz and the Leibniz-Clarke
correspondence. In later times, it can be traced through Mach, Einstein, and
finally to the modern Barbour-Bertotti version. Our contribution is to provide
a quantum version of this historical evolution. It is this history that
extends through several centuries that justifies our use of the name
`Relational Quantum Mechanics' as appropriate. The name has also been adopted
by Rovelli and collaborators in the very different context of quantum
measurement and of correlations among observables. Their line of research has
little to do with the subject of this paper.

Our goal here is to formulate a non-relativistic relational QM within the
framework of Entropic Dynamics (ED) \cite{Caticha 2019}\cite{Caticha 2021}.
Entropic Dynamics is a subject within the subfield of theoretical physics that
is currently called Foundations of Physics. Its goal is to derive or
\textquotedblleft reconstruct\textquotedblright\ the standard formalism of QM
and to resolve the longstanding conceptual problems (wave-particle duality,
the measurement problem, the ontic vs. the epistemic interpretation of the
wave function, wave function collapse, etc.) that have plagued QM since its origin.

The mathematical formalism is derived using well established tools and
principles of inference --- probabilities, entropies, and information
geometry. (For a pedagogical review see \cite{Caticha 2025}.)\ With regard to
interpretation, one appealing feature is that ED achieves a clear separation
between the epistemic and the ontic elements. This allows ED to solve the
measurement problem while evading the no-go theorems that afflict other
realist $\psi$-epistemic interpretations \cite{Caticha 2022}. ED is a
conservative theory in that it grants a definite ontic status to things such
as the positions of particles -- they have definite values at all times -- and
grants a definite epistemic status to probabilities and wave functions without
invoking exotic quantum probabilities. In contrast, ED is radically
non-classical in that there is no ontic dynamics; ED is a purely epistemic
dynamics of probabilities.

As a method of quantization ED is singled out in that it does not rely on the
prior formulation of a classical dynamics to which one must append some ad hoc
quantization rules; therefore there are no operator ordering ambiguities,
because there are no operators. In the present context, one does not first
formulate a classical Machian dynamics \`{a} la Barbour-Bertotti with a
corresponding classical criterion for best matching. Instead one directly
formulates a quantum theory with its corresponding quantum best matching
criterion (BM) based on the natural geometrical tools available to ED, namely,
information and symplectic geometry which lead, eventually, to the Hilbert
space inner product.

Some of Barbour and Bertotti's ideas can be readily adapted and imported into
the ED framework. In an earlier work towards a relational ED \cite{Ipek
Caticha 2016} some of the conceptual issues were successfully settled: in a
dynamics of probabilities it should come as no surprise that the
configurations to be compared involve probability distributions. A second
crucial question is the particular choice of mismatch measure. In the
classical context the measure adopted by Barbour and Bertotti was borrowed
from Jacobi's classical action principle and amounts to a variation on a
least-squares mismatch \cite{Barbour Bertotti 1982}. In \cite{Ipek Caticha
2016} a mismatch measure based on information geometry was adopted which was
natural in the probabilistic context but, unfortunately, that measure proved
less than satisfactory in practice. In the meantime, further formal
developments in the ED reconstruction of QM \cite{Caticha 2021} have shown
that the relevant geometrical structures relevant to QM involve not just
information geometry but also symplectic geometry. Here we deploy the\ recent
development of ED\ as Hamilton-Killing flows and propose a quantum BM
criterion to derive a relational QM of particles.

Our present concern with a non-relativistic relational QM implies that the
resulting theory will necessarily exhibit some still provisional and therefore
unsatisfactory features. Thus, what we produce here are toy models that will
allow us to ask questions and test ideas that will prove useful in the
construction of more realistic relativistic theories.

One such question is what we mean by `relational'. The idea behind absolute
space and time is that these structures exist independently of matter. In
contrast, in a fully relational approach all assertions about space and time
are to be ultimately interpreted as assertions about relations involving
matter and the relevant probabilities and wave functions.

Another question is about the nature of those relations: with respect to what
transformations is the theory supposed to be relational? In a more realistic
model that includes quantum fields and gravity, one might expect the
transformations in question to include local diffeomorphisms. Here, in a
non-relativistic setting, we shall be more modest and include only rigid
translations and rotations. Our models will, therefore, be only partially
relational; they will retain some absolute structures including absolute
simultaneity and the Euclidean geometry of space.

We will argue that as previously formulated ED is already temporally
relational. One goal is to make this feature more explicit and, having done
so, to formulate a non-relativistic quantum model that is generally covariant
with respect to time. This toy model exemplifies a strategy that successfully
evades the notorious \textquotedblleft problem of time\textquotedblright\ in
quantum gravity \cite{Kiefer 2007}\cite{Kuchar 2011}. Briefly, the
\textquotedblleft problem of time\textquotedblright\ arises in the canonical
quantization of gravity\ because in any generally covariant theory the
physical quantum states are constrained to be annihilated by the quantum
Hamiltonian, $\hat{H}\Psi=0$. Thus, the analogue of the Schr\"{o}dinger
equation --- the Wheeler-DeWitt equation --- implies that physical states
cannot evolve and that nothing ever happens.

Here, once again, we wish to avoid the misunderstandings that might arise from
confusion with another approach to dynamics due to Page and Wootters
\cite{Page Wootters 1983} who proposed a \emph{timeless picture of quantum
dynamics}. Their approach to time is instrumentalist: time is what is measured
by a clock. Their universe consists of two systems, one being the system of
interest and the other a clock, and it is assumed to be in a stationary state
--- thus the \emph{timeless} picture. Dynamics consists of tracking the
correlations due to entanglement between the system and the clock. (For a
pedagogical treatment with more recent references, see \cite{Valdes-Hernandez
2020}.) As we shall see below, relational time in the ED approach differs in
several crucial ways. Our universe is not necessarily in a stationary state
and there are no external clocks. Furthermore, as one might expect in an
\emph{entropic} dynamics, there is a natural arrow of time, even though
invariance under time reversal need not be violated.

In section 2, as a preliminary to defining a spatially relational ED, we
establish the subject matter --- the ontic microstates --- and define the
concept of intrinsic change. The ED of intrinsic change is formulated in
section 3 where we emphasize that the notion of entropic time associated to ED
is already relational in the sense that the clock that defines entropic time
is the quantum system itself --- there are neither external clocks nor an
absolute external time. In section 4 we formulate the quantum criterion for
best matching using the natural geometric tools available to ED. In section
4.1 we construct a QM model that is relational with respect to rigid
translations, and then in section 4.2 a QM\ model that is relational with
respect to both rigid translations and rotations. In section 5 we pursue
formal developments: we write the relational QM\ models in terms of a
Hamiltonian action principle. Then, in order to make the temporal
relationality more explicit we rewrite our ED models in a form that is
\textquotedblleft generally covariant\textquotedblright. The procedure is well
known from the canonical formulation of general relativity and results in a
\textquotedblleft parametrized ED\textquotedblright\ \cite{Kuchar
1973}\cite{Henneaux Teitelboim 1992}. Our quantum toy model serves to
illustrate how the ED approach evades the notorious problem of time that
afflicts the canonical quantization of gravity. Finally, in section 6 we
summarize our conclusions.

\section{ Spatial relationality}

\subsection{The ontic microstates}

The system of interest consists of $N$ particles that live in a three
dimensional flat space $\mathbf{X}$ with metric $\delta_{ab}$. The microstate
of the $N$ particle system is a point $x\in\mathbf{X}_{N}$ in the $3N$
dimensional ontic configuration space, $\mathbf{X}_{N}$, with coordinates
$x_{n}^{a}$ ($a=1,2,3$ is the spatial index and $n=1\ldots N$ labels the particle).

The two central assumptions concerning the ontology are that the particles
have definite positions and that they follow continuous paths. (Generic paths
will turn out to be non-differentiable.) Our goal is to predict those
positions on the basis of information that happens to be limited, therefore,
the best we can hope to do is assign a probability distribution $\rho(x)$ and
study its time evolution.\ The continuity of the paths leads to an important
simplification because it implies that the motion can be analyzed as a
sequence of simple infinitesimally short steps. To find the probability
$P(x^{\prime}|x)$ that the system takes a short step from $x_{n}^{a}$ to
$x_{n}^{\prime a}=x_{n}^{a}+\Delta x_{n}^{a}$ we shall use the method of
maximum entropy. Then these short steps will be iterated to find the evolving
$\rho_{t}(x)$. So far this is standard ED as described in \cite{Caticha 2019}
or \cite{Caticha 2025}.

\subsection{Intrinsic change}

For a relational ED we stipulate that the specification of the microstate in
terms of the coordinates $x_{n}^{a}$ is redundant in the sense that shifting
all particles by rigid (or global) translations or rotations, does not lead to
a different ontic state. Relationism about space and time can take different
forms depending on the nature of the transformations being considered.
Consider spatial transformations (displacements) of the form,
\begin{align}
\text{ST1}  &  \text{:\qquad}\vec{x}^{\prime}=R_{0}\,\vec{x}+\vec{\lambda}%
_{0}~,\label{ST1}\\
\text{ST2}  &  \text{:\qquad}\vec{x}^{\prime}=R_{0}\,\vec{x}+\vec{\lambda
}(t)~,\label{ST2}\\
\text{ST3}  &  \text{:\qquad}\vec{x}^{\prime}=R(t)\,\vec{x}+\vec{\lambda}(t)~.
\label{ST3}%
\end{align}
The idea is that with ST1 all particles are subject to rigid displacements
$\vec{\lambda}_{0}$ and rotations $R_{0}$ that are independent of location and
time. Such transformations are not meant to generate a different ontic state
and they preserve the form of the equations of motion. They reflect a weak
form of relationism in that the whole of space-time is rigidly shifted and
rotated. In this paper we shall be concerned with constructing QM models that
reflect the stronger relationism implied by transformations of type ST2 and
the even stronger implied by ST3.

Transformations ST2 allow the greater freedom to displace space by different
rigid amounts $\vec{\lambda}(t)$ at different instants. The corresponding ST2
model\ exhibits a fair relationality with respect to rigid translations but
much less so with respect to rotations. As an aside, the subgroup of ST2
transformations with%
\begin{equation}
\text{ST4:\qquad}\vec{x}^{\prime}=R_{0}\,\vec{x}+\vec{\lambda}_{0}+\vec{v}%
_{0}t~, \label{ST4}%
\end{equation}
where $\vec{v}_{0}$ and $\vec{\lambda}_{0}$ are constant vectors, describes
Galilean relativity. Since the corresponding notion of relative space is what
Newton described in the \emph{Scholium} to the \emph{Principia}, systems that
are relational not with respect to the full ST2 transformations but only with
respect to the more limited subgroup ST4 reflect a \textquotedblleft
Newtonian\textquotedblright\ relationism.

Transformations ST3 allow greater relationality with respect to rotations.
Just as in classical mechanics, transforming to rotating frames of reference
has dynamical consequences; total angular momentum is dynamically relevant.
More on this later.

Let us then consider shifts of the form%

\begin{equation}
x_{n}^{a}\rightarrow x_{n}^{a}+\xi_{n}^{a}\quad\text{where}\quad\xi_{n}%
^{a}(x_{n})=\varepsilon^{abc}\zeta_{b}x_{nc}+\lambda^{a} \label{shift a}%
\end{equation}
($\lambda^{a}$ and $\zeta^{a}$ are some arbitrary vectors independent of the
particle $n$ and of the position $x_{n}^{a}$ and $\varepsilon^{abc}$ is the
Levi-Civita tensor). Since $x_{n}^{a}$ and $x_{n}^{a}+\xi_{n}^{a}$ represent
the same initial state and $x_{n}^{\prime a}$ and $x_{n}^{\prime a}+\xi
_{n}^{\prime a}$ represent the same final state, the short step from one to
the other will be represented by
\begin{equation}
(x_{n}^{\prime a}+\xi_{n}^{\prime a})-(x_{n}^{a}+\xi_{n}^{a})=\Delta x_{n}%
^{a}+\Delta\xi_{n}^{a}~,
\end{equation}
The two configurations are said to be \textquotedblleft best
matched\textquotedblright\ when the two vectors $\Delta\lambda^{a}$ and
$\Delta\zeta^{a}$ are chosen to minimize a certain measure of mismatch to be
defined later. Finding the optimal $\Delta\lambda_{\text{best}}^{a}$ and
$\Delta\zeta_{\text{best}}^{a}$ amounts to deciding which position $x^{\prime
a}$ at the later instant \emph{is the same as}\ the position $x^{a}$ at the
earlier instant. It provides a criterion of \textquotedblleft
equilocality\textquotedblright\ between successive instants. The quantity
\begin{equation}
\hat{\Delta}x_{n}^{a}=\Delta x_{n}^{a}+\Delta\xi_{n\,\text{best}}^{a}
\label{Delta x}%
\end{equation}
will be called the \emph{intrinsic change}. Two successive configurations are
\textquotedblleft intrinsically identical\textquotedblright\ when equilocality
has been established and $\hat{\Delta}x_{n}^{a}=0$. At this point in our
argument the optimal $\Delta\lambda_{\text{best}}^{a}$ and $\Delta
\zeta_{\text{best}}^{a}$ are not yet known. However, to proceed we will assume
that equilocality\ has been established through some trial shift $\Delta
\xi_{n\,\text{best}}^{a}$ to be determined later. From now on we drop the
subscript `best' and write $\Delta\xi_{n\,\text{best}}^{a}=\Delta\xi_{n}^{a}$.

\section{The entropic dynamics of intrinsic change}

\label{ED from entropy}Except for the replacement of the change $\Delta
x_{n}^{a}$ by the intrinsic change $\hat{\Delta}x_{n}^{a}$ the contents of
this section parallels closely the material described in \cite{Caticha 2019}
or in Chapter 11 of \cite{Caticha 2025}.

\subsection{The transition probability for a short step}

To find the transition probability we maximize the entropy of $P\left(
x^{\prime}|x\right)  $ relative to a prior $Q\left(  x^{\prime}|x\right)  $,%
\begin{equation}
S[P,Q]=-\int dx^{\prime}P\left(  x^{\prime}|x\right)  \log\frac{P\left(
x^{\prime}|x\right)  }{Q\left(  x^{\prime}|x\right)  }\ , \label{entropy a}%
\end{equation}
subject to constraints that codify the relevant physical information. We
choose a prior $Q(x^{\prime}|x)$ that describes a state of knowledge that is
common to all short steps \emph{before} we take into account the additional
information that is specific to each specific short step. We require it to
incorporate the information that the particles take infinitesimally short
steps but $Q$ is otherwise maximally uninformative in the sense that it must
reflect the translational and rotational invariance of the space $\mathbf{X}$
and expresses total ignorance about any correlations. (Such a prior can itself
be derived from the principle of maximum entropy.) The chosen prior is
\begin{equation}
Q(x^{\prime}|x)\propto\exp\left(  -\frac{1}{2}%
%TCIMACRO{\dsum \nolimits_{n}}%
%BeginExpansion
{\displaystyle\sum\nolimits_{n}}
%EndExpansion
\alpha_{n}\delta_{ab}\hat{\Delta}x_{n}^{a}\hat{\Delta}x_{n}^{b}\right)  ~.
\label{prior}%
\end{equation}
The multipliers $\alpha_{n}$ are constants that may depend on the index $n$ in
order to describe non-identical particles. To enforce the fact that the steps
are meant to be infinitesimally short, the $\alpha_{n}$s will eventually be
taken to infinity.

The information that induces the directionality and correlations specific to
each individual short step is introduced by imposing one additional
constraint,
\begin{equation}
\sum_{n}\left\langle \hat{\Delta}x_{n}^{a}\right\rangle \frac{\partial
\varphi\left(  x\right)  }{\partial x_{n}^{a}}=\int dx^{\prime}P\left(
x^{\prime}|x\right)  \sum_{n}\hat{\Delta}x_{n}^{a}\frac{\partial\varphi\left(
x\right)  }{\partial x_{n}^{a}}=\kappa^{\prime}. \label{Phase constraint}%
\end{equation}
The function $\varphi(x)=\varphi(x_{1},\ldots,x_{N})$, called the
\textquotedblleft drift\textquotedblright\ potential, will play a central role
in what follows. The quantity $\kappa^{\prime}$ is a small but for now
unspecified constant. Maximizing the entropy (\ref{entropy a}) subject to the
constraints (\ref{Phase constraint}) and normalization leads to a Gaussian
distribution,%
\begin{equation}
P\left(  x^{\prime}|x\right)  =\frac{1}{Z}\exp\left[  -%
%TCIMACRO{\dsum \nolimits_{n}}%
%BeginExpansion
{\displaystyle\sum\nolimits_{n}}
%EndExpansion
\frac{\alpha_{n}}{2}\delta_{ab}\left(  \hat{\Delta}x_{n}^{a}-\frac
{\alpha^{\prime}}{\alpha_{n}}\partial_{n}^{a}\varphi\right)  \left(
\hat{\Delta}x_{n}^{b}-\frac{\alpha^{\prime}}{\alpha_{n}}\partial_{n}%
^{b}\varphi\right)  \right]  . \label{TransProb}%
\end{equation}
where $Z$ is a normalization constant, $\left\{  \alpha_{n},\alpha^{\prime
}\right\}  $ are Lagrange multipliers, and $\partial_{na}=\partial/\partial
x_{n}^{a}$. Substituting (\ref{Delta x}) into (\ref{TransProb}), a generic
displacement $\Delta x_{n}^{a}=x_{n}^{\prime a}-x_{n}^{a}$ can then be written
as the sum of an expected drift plus a fluctuation,
\begin{equation}
\Delta x_{n}^{a}=\langle\Delta x_{n}^{a}\rangle+\Delta w_{n}^{a}\,\,,
\label{delta x a}%
\end{equation}
where
\begin{equation}
\langle\hat{\Delta}x_{n}^{a}\rangle=\frac{\alpha^{\prime}}{\alpha_{n}}%
\delta^{ab}\partial_{nb}\varphi~,\quad\langle\Delta x_{n}^{a}\rangle
=\frac{\alpha^{\prime}}{\alpha_{n}}\delta^{ab}\partial_{nb}\varphi-\Delta
\xi_{n}^{a}~, \label{drift a}%
\end{equation}%
\begin{equation}
\left\langle \Delta w_{n}^{a}\right\rangle =\left\langle \hat{\Delta}x_{n}%
^{a}-\frac{\alpha^{\prime}}{\alpha_{n}}\partial_{n}^{a}\varphi\right\rangle
=0~,\quad\text{and}\quad\langle\Delta w_{n}^{a}\Delta w_{n^{\prime}}%
^{b}\rangle=\frac{1}{\alpha_{n}}\delta^{ab}\delta_{nn^{\prime}}~.
\label{fluct a}%
\end{equation}

\subsection{Temporal relationality: entropic time}

An epistemic dynamics of probabilities demands an epistemic notion of time.
The construction of time involves introducing the concept of an instant,
verifying that the instants are suitably ordered, and adopting a convenient
definition of duration, that is, a measure of the interval between instants.
The construction is intimately related to information and inference.

An instant is an epistemically complete state specified by information ---
codified into the functions $\rho_{t}(x)$ and $\varphi_{t}(x)$ --- that is
sufficient for generating the next instant. Thus, the instant we call the
present is defined so that,\ given the information codified into the present
instant, the future is independent of the past. Formally, ED is a Markovian
process: if the distribution $\rho_{t}(x)$ and the drift potential
$\varphi_{t}(x)$ refer to the instant $t$, then the distribution
\begin{equation}
\rho_{t\prime}\left(  x^{\prime}\right)  =\int dx\,P\left(  x^{\prime
}|x\right)  \rho_{t}\left(  x\right)  \label{DefTime}%
\end{equation}
generated by $\rho_{t}(x)$ and $P\left(  x^{\prime}|x\right)  $ defines what
we mean by the \textquotedblleft next\textquotedblright\ instant $t^{\prime}$.
(This takes care of the equation of evolution for $\rho_{t}(x)$; below we will
return to specify the evolution of $\varphi_{t}(x)$.)

In ED time is \emph{constructed} instant by instant and since no reference is
made to external clocks the dynamics does not unfold in a pre-existing
externally given absolute time or space-time. Furthermore, the construction
leads to a sequence of instants that are ordered because the transition
probability $P\left(  x^{\prime}|x\right)  $ is determined by
\emph{maximizing} an entropy and, by its very construction, there is a natural
arrow of time.

The last ingredient -- duration, the interval $\Delta t$ between successive
instants -- is defined to simplify the dynamics. The description is simplest
when it reflects the symmetry of translations in space and time that are
typical of the weak interactions that characterize non-relativistic physics.
In Newtonian mechanics the prototype of a clock is a free particle that moves
equal distances in equal times. In ED the dynamics is described by $P\left(
x^{\prime}|x\right)  $ and we define duration so that the multipliers
$\alpha_{n}$ and $\alpha^{\prime}$ are constants independent of $x$ and $t$ so
they lead to an entropic time that resembles a (relational) Newtonian time in
that it flows \textquotedblleft equably everywhere and
everywhen.\textquotedblright\ Here too we define duration so that for
sufficiently short steps there is a well defined drift velocity. As we see
from eq.(\ref{drift a}) this is achieved by setting the ratio $\alpha^{\prime
}/\alpha_{n}\ $proportional to $\Delta t$. Thus, \emph{the transition
probability provides us with a clock}. For future convenience the
proportionality constants will be expressed in terms of some particle-specific
constants $m_{n}$,
\begin{equation}
\frac{\alpha^{\prime}}{\alpha_{n}}=\frac{1}{m_{n}}\Delta t~.
\label{alpha ratio a}%
\end{equation}
At this point the constants $m_{n}$ receive no interpretation beyond the fact
that their dependence on the particle label $n$ recognizes that the particles
need not be identical but later we shall see that the $m_{n}$s will be
identified with the particle masses. To explore the consequences of the choice
(\ref{alpha ratio a}) we assume $\alpha^{\prime}=$ $1/\eta$ is a constant and
choose $\eta$ so that if $\Delta t$ has units of time, then $m_{n}$ has units
of mass. Then,
\begin{equation}
\alpha^{\prime}=\frac{1}{\eta}\quad\text{so that\quad}\alpha_{n}=\frac{m_{n}%
}{\eta\Delta t}~.
\end{equation}

We rewrite the generic short step eqs.(\ref{delta x a})-(\ref{fluct a}) in a
more streamlined notation,
\begin{equation}
\Delta x^{A}=\hat{\Delta}x^{A}-\Delta\xi^{A}=\left\langle \Delta
x^{A}\right\rangle +\Delta w^{A},
\end{equation}
where $A=(n,a)$ is a composite index that includes both the particle $n$ and
the spatial index $a$ so that $x_{n}^{a}=x^{A}$. To simplify we write the
spatial shift in configuration space as
\begin{equation}
\xi^{A}(x)=\xi_{n}^{a}(x_{n})~.
\end{equation}
We also introduce the mass tensor,
\begin{equation}
m_{AB}=m_{n}\delta_{AB}=m_{n}\delta_{nn^{\prime}}\delta_{ab}\quad\text{and its
inverse}\quad m^{AB}=\frac{1}{m_{n}}\delta^{AB}~.
\end{equation}
Then we find that the shift $\Delta\xi^{A}$ affects the expected steps,
\begin{equation}
\left\langle \Delta x^{A}\right\rangle =\langle\hat{\Delta}x^{A}%
\rangle-\langle\Delta\xi^{A}\rangle=\Delta t\,m^{AB}\partial_{B}\varphi
-\Delta\xi^{A}\ , \label{ExpX}%
\end{equation}
but, of course, does not affect the fluctuations,
\begin{equation}
\left\langle \Delta x^{A}\Delta x^{B}\right\rangle =\left\langle \hat{\Delta
}x^{A}\hat{\Delta}x^{B}\right\rangle +O(\Delta t^{2})~.
\end{equation}
Therefore,
\begin{equation}
\hat{\Delta}w^{A}=\Delta w^{A}\quad\text{with}\quad\left\langle \Delta
w^{A}\right\rangle =0\quad\text{and\quad}\left\langle \Delta w^{A}\Delta
w^{B}\right\rangle =\eta\Delta t\,m^{AB}, \label{ExpW}%
\end{equation}
which remain large $\Delta w\sim O(\Delta t^{1/2})$ and essentially
isotropic.\ This leads us to expect that the optimal shift $\Delta\xi$ is of
order $\Delta t$ so that $\Delta\xi\ll\Delta w$.

We conclude this section noting that the entropic time we have constructed
here is already fully relational. To summarize, time is the sequence of
instants that are epistemically complete: each instant is defined by
information codified into the distributions $\rho_{t}(x)$ and $\varphi_{t}(x)$
that serve to define the next instant. The evolution equation (\ref{DefTime})
is used both to define the dynamics and to construct time itself instant by
instant. There is neither an absolute time nor are there external clocks. As
we can see from (\ref{ExpX}) or (\ref{ExpW}) the system is its own clock in
that the duration $\Delta t$ can be defined either from the expected intrinsic
change $\langle\hat{\Delta}x^{A}\rangle$ or from the fluctuations
$\left\langle \Delta w^{A}\Delta w^{B}\right\rangle $.

\subsection{The probability evolution equation}

The dynamical equation of evolution, eq.(\ref{DefTime}) can be written in
differential form as a continuity equation (a Fokker-Planck equation),
\begin{equation}
\partial_{t}\rho\left(  x,t\right)  =-\partial_{A}\left[  \rho\left(
x,t\right)  v^{A}\left(  x,t\right)  \right]  ~, \label{FP a}%
\end{equation}
where $v^{A}$ is the velocity of the probability flow, or current velocity,%
\begin{equation}
v^{A}\left(  x,t\right)  =m^{AB}\partial_{B}\phi\left(  x,t\right)  -\dot{\xi
}^{A}\text{\quad with\quad}\phi=\varphi-\eta\log\rho^{1/2}~,
\label{CurrentVel}%
\end{equation}
and $\dot{\xi}^{A}=\Delta\xi^{A}/\Delta t$. The derivation, which involves a
technique that is well known from diffusion theory \cite{Chandrasekhar 1943},
follows the same steps as the analogous (non-relational) derivation found in
\cite{Caticha 2025}.

It turns out that the crucial drift potential $\varphi$ that was introduced
via the maximum entropy constraint (\ref{Phase constraint}), will always
appear in the particular combination of $\varphi$ and $\log\rho$ given in
(\ref{CurrentVel}). This justifies introducing the function $\phi(x,t)$ that
will play three separate roles: first, as we saw above, it is related to a
constraint in the maximization of entropy; second, if the probabilities
$\rho(x,t)$ are considered as generalized coordinates, then $\phi(x,t)$ turns
out to be a convenient choice for the momenta that are canonically conjugate
to them; and third, $\phi(x,t)$ will turn out to be the phase of the quantum
wave function, $\psi=\rho^{1/2}e^{i\phi/\hbar}$.

The current velocity $v^{A}$ receives three types of contributions. The first
two are the familiar drift and osmotic velocities described through the
gradient of the \textquotedblleft phase\textquotedblright\ $\phi$
\cite{Caticha 2025}. The third contribution is the term that implements
relationality, the shift velocity $\dot{\xi}^{A}$.

We now return to the continuity equation (\ref{FP a}) and rewrite it in an
alternative form. The important observation is that a functional $\tilde
{H}[\rho,\phi]$ can be found such that (\ref{FP a}) can be written as
\begin{equation}
\partial_{t}\rho_{t}(x)=\frac{\delta\tilde{H}}{\delta\phi(x)}%
~.\label{Hamilton a}%
\end{equation}
The desired $\tilde{H}$ satisfies
\begin{equation}
-\partial_{A}\left[  \rho_{t}m^{AB}(\partial_{B}\phi-\dot{\xi}_{B})\right]
=\frac{\delta\tilde{H}}{\delta\phi(x)}\quad\text{where}\quad\dot{\xi}%
_{B}=m_{BC}\dot{\xi}^{C}~,
\end{equation}
which is a linear functional equation that can be easily integrated. The
result is%
\begin{equation}
\tilde{H}[\rho,\phi]=\int dx\,\frac{1}{2}\rho m^{AB}(\partial_{A}\phi-\dot
{\xi}_{A})(\partial_{B}\phi-\dot{\xi}_{B})+\tilde{F}[\rho
]~,\label{Hamiltonian a}%
\end{equation}
where the unspecified functional $\tilde{F}[\rho]$ is an integration constant
(\emph{i.e.}, independent of $\phi$). This maneuver has allowed us to identify
the kinetic part of the ED Hamiltonian and a \textquotedblleft
potential\textquotedblright\ $\tilde{F}$ that is independent of $\phi$ but
could potentially also depend on $t$. More importantly, it suggests a
Hamiltonian framework in which the epistemic variables $\rho$ and $\phi$ are
canonically conjugate. The epistemic configuration space (or \textquotedblleft
e-configuration\textquotedblright\ space) is the simplex,
\begin{equation}
\mathcal{S}=\left\{  \rho|\text{~}\rho(x)\geq0\,;~\int dx\,\rho(x)=1\right\}
~.
\end{equation}
The epistemic phase space (or \textquotedblleft e-phase\textquotedblright%
\ space) is the cotangent bundle $T^{\ast}\mathcal{S}$ \cite{Caticha 2025}. As
is well known, dealing with normalized probabilities is a technical
inconvenience that can be handled by discarding the normalization constraint
and dealing instead with the larger space
\begin{equation}
\mathcal{S}^{+}=\left\{  \rho|\text{~}\rho(x)\geq0\right\}
\end{equation}
of unnormalized probabilities and its associated cotangent bundle $T^{\ast
}\mathcal{S}^{+}$.

\subsection{The symplectic, metric, and complex structures of e-phase space}

For details of the constructions in this and the next sections see
\cite{Caticha 2021}\cite{Caticha 2025}. First we shall establish some
notation. To simplify we shall write $\rho(x)=\rho_{x}$ and $\phi(x)=\phi_{x}%
$. A point $(\rho,\phi)$ in e-phase space $T^{\ast}\mathcal{S}^{+}$ has
coordinates $(\rho_{x},\phi_{x})$. A curve in e-phase space parametrized by
$\lambda$ is the one-dimensional set of points\ $(\rho(\lambda),\phi
(\lambda))$. The vector $\bar{V}$ tangent to the curve is written
\begin{equation}
\bar{V}=\frac{d}{d\lambda}=\int dx\left(  \frac{d\rho_{x}}{d\lambda}%
\frac{\delta}{\delta\rho_{x}}+\frac{d\phi_{x}}{d\lambda}\frac{\delta}%
{\delta\phi_{x}}\right)  =V^{\alpha x}\frac{\delta}{\delta X^{\alpha x}}%
\end{equation}
where $dx=d^{3N}x$, we introduced the discrete index $\alpha=1,2$ to stand for
$\rho$ and $\phi$ respectively,
\begin{equation}
X^{1x}=\rho_{x}\quad\text{and}\quad X^{2x}=\phi_{x}~,\quad
\end{equation}
and we adopt the convention of integration over repeated indices. The
directional derivative of a functional $F[\rho,\phi]$ along $\bar{V}$ is%
\begin{equation}
\frac{dF}{d\lambda}=\int dx\left(  \frac{\delta F}{\delta\rho_{x}}\frac
{d\rho_{x}}{d\lambda}+\frac{\delta F}{\delta\phi_{x}}\frac{d\phi_{x}}%
{d\lambda}\right)  =\frac{\delta F}{\delta X^{\alpha x}}\frac{dX^{\alpha x}%
}{d\lambda}=\tilde{\nabla}F[\bar{V}]~,
\end{equation}
where $\tilde{\nabla}$ is the gradient in $T^{\ast}\mathcal{S}^{+}$, that is,
\begin{equation}
\tilde{\nabla}F=\int dx\left(  \frac{\delta F}{\delta\rho_{x}}\tilde{\nabla
}\rho_{x}+\frac{\delta F}{\delta\phi_{x}}\tilde{\nabla}\phi_{x}\right)
=\frac{\delta F}{\delta X^{\alpha x}}\tilde{\nabla}X^{\alpha x}~,
\end{equation}
where
\begin{equation}
\tilde{\nabla}X^{1x}=\tilde{\nabla}\rho_{x}\quad\text{and}\quad\tilde{\nabla
}X^{2x}=\tilde{\nabla}\phi_{x}%
\end{equation}
are the basis covectors. (The tilde `\symbol{126}' serves to distinguish the
gradient $\nabla$ on $\mathcal{S}^{+}$\ from the gradient $\tilde{\nabla}$ on
$T^{\ast}\mathcal{S}^{+}$.)

Once local coordinates $\rho_{x}$ and $\phi_{x}$ on the e-phase space have
been identified there is a natural symplectic form%

\begin{equation}
\Omega\lbrack\cdot,\cdot]=\int dx\,\left(  \tilde{\nabla}\rho_{x}%
[\cdot]\otimes\tilde{\nabla}\phi_{x}[\cdot]-\tilde{\nabla}\phi_{x}%
[\cdot]\otimes\tilde{\nabla}\rho_{x}[\cdot]\right)  ~, \label{x sympl form a}%
\end{equation}
where $\otimes$ is the tensor product. The action of $\Omega\lbrack\cdot
,\cdot]$ on two vectors $\bar{V}=d/d\lambda$ and $\bar{U}=d/d\mu$ is given by
\begin{equation}
\Omega\lbrack\bar{V},\bar{U}]=\int dx\,\left[  V^{1x}U^{2x}-V^{2x}%
U^{1x}\right]  =\Omega_{\alpha x,\beta x^{\prime}}V^{\alpha x}U^{\beta
x^{\prime}}~,~ \label{x sympl form b}%
\end{equation}
where the components of $\Omega$ displayed as a matrix are
\begin{equation}
\lbrack\Omega_{\alpha x,\beta x^{\prime}}]=%
\begin{bmatrix}
0 & 1\\
-1 & 0
\end{bmatrix}
\delta_{xx^{\prime}}~, \label{x sympl form c}%
\end{equation}
and $\delta_{xx^{\prime}}=\delta(x,x^{\prime})$ is the Dirac $\delta$ function.

Time evolution is required to reproduce the continuity equation. It will also
preserve the normalization constraint,
\begin{equation}
\tilde{N}=0\quad\text{where}\quad\tilde{N}=1-\left\vert \rho\right\vert
\quad\text{and}\quad\left\vert \rho\right\vert \overset{\text{def}}{=}\int
dx\,\rho_{x}~. \label{x N constraint}%
\end{equation}
Indeed, one can check that
\begin{equation}
\partial_{t}\tilde{N}=\{\tilde{N},\tilde{H}\}=0~. \label{x N conservation}%
\end{equation}

The embedding e-configuration space $\mathcal{S}^{+}$ and e-phase space
$T^{\ast}\mathcal{S}^{+}$ are (without loss of generality \cite{Caticha 2025})
assigned the simplest geometries consistent with information geometry, namely,
they are flat. With this choice the $T^{\ast}\mathcal{S}^{+}$ metric is
\begin{equation}
\delta\tilde{\ell}^{2}=\int dx\left[  \frac{\hbar}{2\rho_{x}}(\delta\rho
_{x})^{2}+\frac{2\rho_{x}}{\hbar}\delta\phi_{x}^{2}\right]  =G_{\alpha x,\beta
x^{\prime}}\delta X^{\alpha x}\delta X^{\beta x^{\prime}}\label{TS+ metric c}%
\end{equation}
where the metric tensor $G$ displayed as a matrix is
\begin{equation}
\lbrack G_{\alpha x,\beta x^{\prime}}]=%
\begin{bmatrix}
\frac{\hbar}{2\rho_{x}}\delta_{xx^{\prime}} & 0\\
0 & \frac{2\rho_{x}}{\hbar}\delta_{xx^{\prime}}%
\end{bmatrix}
\,,\label{TS+ metric d}%
\end{equation}
and $\hbar$ is just an arbitrary constant that will eventually be identified
with Planck's constant $h/2\pi$. It is interesting, however, that in ED the
role of $\hbar$ can be characterized geometrically. The constant $\hbar$
determines the relative weights with which the coordinate components
$\delta\rho_{x}$ and the momentum components $\delta\phi_{x}$ contribute to
the length of a vector $(\delta\rho,\delta\phi)$.

The joint existence of symplectic $\Omega$ and metric $G$ structures implies
the existence of a complex structure described by a tensor $J$ defined by
\begin{equation}
J^{\alpha x}{}_{\beta x^{\prime}}=-G^{\alpha x,\gamma x^{\prime\prime}}%
\Omega_{\gamma x^{\prime\prime},\beta x^{\prime}}\quad\text{or}\quad\lbrack
J^{x}{}_{x^{\prime}}]=%
\begin{bmatrix}
0 & -\frac{2\rho_{x}}{\hbar}\delta_{xx^{\prime}}\\
\frac{\hbar}{2\rho_{x}}\delta_{xx^{\prime}} & 0
\end{bmatrix}
~. \label{J tensor}%
\end{equation}
This suggests a canonical transformation to complex coordinates,
\begin{equation}
\psi_{x}=\rho_{x}^{1/2}e^{i\phi_{x}/\hbar}~, \label{psi coords}%
\end{equation}
with conjugate momenta $i\hbar\psi_{x}^{\ast}$ that are naturally adapted to
the complex structure (see \cite{Caticha 2021}\cite{Caticha 2025}). In wave
function coordinates eqs.(\ref{TS+ metric c}) and (\ref{TS+ metric d}) become
\begin{equation}
\delta\tilde{\ell}^{2}=2\hbar\int dx\,\delta\psi_{x}^{\ast}\delta\psi
_{x}=2\hbar\langle\delta\psi|\delta\psi\rangle~,\quad\lbrack G_{xx^{\prime}%
}]=-i\delta_{xx^{\prime}}%
\begin{bmatrix}
0 & 1\\
1 & 0
\end{bmatrix}
~. \label{TS+ metric e}%
\end{equation}
Since the transformation (\ref{psi coords}) is canonical, the components of
the symplectic tensor $\Omega\,$, eq.(\ref{x sympl form c}), remain unchanged.

\subsection{Hamilton-Killing flows}

We seek an Entropic Dynamics that preserves the natural geometric structures
of the e-phase space. This requires a Hamiltonian $\tilde{H}$ that
simultaneously preserves the symplectic structure $\Omega$ and the metric
structure $G$. In order to generate flows that are simultaneously Hamilton
flows and Killing flows $\tilde{H}$ must satisfy
\begin{equation}
\pounds _{H}\Omega=0\quad\text{and}\quad\pounds _{H}G=0~, \label{HK flows}%
\end{equation}
where $\pounds _{H}$ is the Lie derivative along the Hamiltonian vector field
$\bar{H}$. As shown in \cite{Caticha 2021}\cite{Caticha 2025} the
Hamilton-Killing flows (or HK flows) that also preserve normalization are
generated by Hamiltonian functionals that are bilinear in $\psi_{x}^{\ast}$
and $\psi_{x}$,
\begin{equation}
\tilde{H}[\psi,\psi^{\ast}]=\int dxdx^{\prime}\,\psi_{x}^{\ast}\hat
{H}_{xx^{\prime}}\psi_{x^{\prime}}~, \label{bilinear hamiltonian b}%
\end{equation}
which, incidentally, implies the linearity of the Schr\"{o}dinger equation. In
addition we require that $\tilde{H}$ generate evolution in entropic time, that
is, it must agree with eq.(\ref{Hamiltonian a}) in order to reproduce the
continuity equation (\ref{Hamilton a}). This constrains the integration
constant $\tilde{F}[\rho]$ to a bilinear functional with a potential $V(x)$
that is local in configuration space. The resulting Hamiltonian is
\begin{equation}
\tilde{H}_{\dot{\xi}}[\psi,\psi^{\ast}]=\int dx\psi_{x}^{\ast}%
%TCIMACRO{\tsum \limits_{n}}%
%BeginExpansion
{\textstyle\sum\limits_{n}}
%EndExpansion
\left(  \frac{1}{2m_{n}}\delta^{ab}(\frac{\hbar}{i}\partial_{na}-m_{n}\dot
{\xi}_{na})(\frac{\hbar}{i}\partial_{nb}-m_{n}\dot{\xi}_{nb})+V(x)\right)
\psi_{x}~, \label{Hamiltonian c}%
\end{equation}
where we used%
\begin{equation}
\dot{\xi}_{A}=m_{AC}\dot{\xi}^{C}=%
%TCIMACRO{\tsum \limits_{n^{\prime}}}%
%BeginExpansion
{\textstyle\sum\limits_{n^{\prime}}}
%EndExpansion
m_{n}\delta_{nn^{\prime}}\delta_{ac}\dot{\xi}_{n}^{c}=m_{n}\delta_{ac}\dot
{\xi}_{n}^{c}=m_{n}\dot{\xi}_{na}~.
\end{equation}
This bilinear Hamiltonian leads to a linear Schr\"{o}dinger equation,
\begin{equation}
i\hbar\partial_{t}\psi=%
%TCIMACRO{\tsum \limits_{n}}%
%BeginExpansion
{\textstyle\sum\limits_{n}}
%EndExpansion
\frac{1}{2m_{n}}\delta^{ab}(\frac{\hbar}{i}\partial_{na}-m_{n}\dot{\xi}%
_{na})(\frac{\hbar}{i}\partial_{nb}-m_{n}\dot{\xi}_{nb})\psi+V(x)\psi~.
\label{SchroEqn}%
\end{equation}

Naturally, in order that linear and angular momentum be conserved, we shall
further demand that the transformations ST1, eq.(\ref{ST1}), which are a
subgroup of ST2 and ST3, be a symmetry of the system. To achieve this we
require that the potential $V(x)$ be a function of the interparticle
distances,
\begin{equation}
V(x)=V(\{|\vec{x}_{n}-\vec{x}_{m}|\})~.
\end{equation}

\section{A quantum criterion for best matching}

There are two foundational pillars of the ED approach to QM: one refers to
kinematics, the other to dynamics. The former is the kinematic concept of
Hamilton-Killing flows, eq.(\ref{HK flows}), based on the information metric
and symplectic tensors. The latter is the explanation of the dynamics of
probabilities as a form of entropic updating described in section
\ref{ED from entropy}. It is only natural to expect that, in its ED version,
the logic of quantum best matching (BM) will also involve the same two tensors
$G$ and $\Omega$. Consider two states $\Psi$ and $X$; in wave function
coordinates,
\begin{equation}
\Psi^{\mu x}=\binom{\psi_{x}}{i\hbar\psi_{x}^{\ast}}\quad\text{and}\quad
X^{\nu y}=\binom{\chi_{y}}{i\hbar\chi_{y}^{\ast}}~,
\end{equation}
where the discrete index $\mu=1,2$ stands for $\psi$ and its momentum
$i\hbar\psi^{\ast}$ respectively. From $\Omega$ and $G$,
eqs.(\ref{x sympl form c}) and (\ref{TS+ metric e}), we define the inner
product (see \cite{Caticha 2021} or \cite{Caticha 2025}) of $\psi_{x}$ and
$\chi_{x}$ by
\begin{equation}
\frac{1}{2\hbar}\int dxdy\,\left(  G_{\mu x,\nu y}+i\Omega_{\mu x,\nu
y}\right)  \Psi^{\mu x}X^{\nu y}=\int dx\,\psi_{x}^{\ast}\chi_{x}%
\overset{\text{def}}{=}\langle\psi|\chi\rangle~.\label{inner prod a}%
\end{equation}

Our goal is to find the shift $x_{n}^{a}\rightarrow x_{n}^{a}+\xi_{n}^{a}$ so
that the state $\psi_{t+dt}$ is best matched relative to the slightly earlier
state $\psi_{t}$. The strategy is to minimize an appropriate measure of the
mismatch between $\psi_{t}$ and $\psi_{t+dt}$. We assume that the states
$\psi_{t}$ and $\psi_{t+dt}$ are normalized and that $\psi_{t+dt}$ is obtained
by time-evolving $\psi_{t}$ using the Hamiltonian $\tilde{H}_{\dot{\xi}}$,
eqs.(\ref{bilinear hamiltonian b}) and (\ref{Hamiltonian c}).\ We propose the
following candidate for a measure of mismatch,%

\begin{equation}
\delta(\dot{\xi})=|\langle\psi_{t}|\psi_{t+dt}\rangle-1|^{2}~. \label{BM a}%
\end{equation}
Then the BM shift $\dot{\xi}_{\text{best}}$ is found by minimizing the measure
$\delta(\dot{\xi})$. Other mismatch measures are in principle possible, but
$\delta(\dot{\xi})$ recommends itself because \textbf{(a)} it involves
geometric structures ($\Omega$ and $G$) that are natural to QM, \textbf{(b)}
it obeys the natural limiting condition of vanishing as $dt\rightarrow0$ so
that $\psi_{t}$ is best-matched relative to itself, \textbf{(c)} one can check
that $\delta(\dot{\xi})$ does have a minimum and, most importantly,
\textbf{(d)} $\delta(\dot{\xi})$ is useful and convenient in actual practice.
Equation (\ref{BM a}) can be written in a more convenient form using
\begin{equation}
|\psi_{t+dt}\rangle=|\psi_{t}\rangle+|\delta\psi_{t}\rangle=|\psi_{t}%
\rangle+\frac{dt}{i\hbar}\hat{H}_{\dot{\xi}}|\psi_{t}\rangle~.
\end{equation}
Then we find
\begin{equation}
\delta(\dot{\xi})=|\langle\psi_{t}|\delta\psi_{t}\rangle|^{2}=\frac{dt^{2}%
}{\hbar^{2}}\langle\psi_{t}|\hat{H}_{\dot{\xi}}|\psi_{t}\rangle^{2}~,
\end{equation}
or,
\begin{equation}
\delta(\dot{\xi})=\frac{dt^{2}}{\hbar^{2}}\tilde{H}_{\dot{\xi}}^{2}~.
\label{BM b}%
\end{equation}
Next we analyze the consequences of minimizing the mismatch $\delta(\dot{\xi
})$.

\subsection{Best matching under rigid translations}

In this section we formulate a QM model that is relational with respect to
rigid translations. Minimizing (\ref{BM b}) with $\tilde{H}_{\dot{\xi}}$ given
by (\ref{Hamiltonian c}) and $\dot{\xi}^{a}=\dot{\lambda}^{a}$,
\begin{equation}
\frac{\partial}{\partial\dot{\lambda}^{a}}\tilde{H}_{\dot{\lambda}}=0~,
\label{BM c}%
\end{equation}
yields%

\begin{equation}
0=-\int dx\psi^{\ast}%
%TCIMACRO{\tsum \limits_{n}}%
%BeginExpansion
{\textstyle\sum\limits_{n}}
%EndExpansion
\delta^{ab}(\frac{\hbar}{i}\partial_{nb}-m_{n}\dot{\lambda}_{b})\psi
=M\dot{\lambda}^{a}-\delta^{ab}\int dx\psi^{\ast}%
%TCIMACRO{\tsum \limits_{n}}%
%BeginExpansion
{\textstyle\sum\limits_{n}}
%EndExpansion
\frac{\hbar}{i}\partial_{na}\psi~,
\end{equation}
where $M=%
%TCIMACRO{\tsum \nolimits_{n}}%
%BeginExpansion
{\textstyle\sum\nolimits_{n}}
%EndExpansion
m_{n}$. Therefore, the optimal shift is given by
\begin{equation}
M\dot{\lambda}_{a}=\langle\psi|\hat{P}_{a}|\psi\rangle=\tilde{P}_{a}~,
\end{equation}
where the (expected) total momentum,
\begin{equation}
\tilde{P}_{a}[\psi,\psi^{\ast}]=\int dx\psi^{\ast}%
%TCIMACRO{\tsum \limits_{n}}%
%BeginExpansion
{\textstyle\sum\limits_{n}}
%EndExpansion
\frac{\hbar}{i}\partial_{na}\psi~, \label{P total}%
\end{equation}
is the generator of global translations. The interpretation is
straightforward. If we are given a sequence of consecutive states, the shift
velocity $\dot{\lambda}^{a}$ that achieves equilocality is given by the
expected velocity of the center of mass.

Conversely, to formulate a relational dynamics with evolution given by
(\ref{Hamiltonian c}) with $\dot{\lambda}^{a}=\dot{\xi}^{a}$ chosen to achieve
best matching, then we must impose that $\dot{\lambda}^{a}$ be a constant in
time and require that solutions be constrained to the subspace of the full
Hilbert space with a given total momentum,
\begin{equation}
\tilde{P}_{a}[\psi,\psi^{\ast}]-M\dot{\lambda}_{a}=0\quad\text{or}\quad
\langle\psi|(\hat{P}-M\dot{\lambda}^{a})|\psi\rangle=0~. \label{BM constr a}%
\end{equation}
This completes our formulation of an ED model that is relational relative to
the rigid translations ST2, but\ we can can go a bit further.

The Galilean transformations ST4, eq.(\ref{ST4}), allow us to describe the
system in the center of mass frame, that is, $\dot{\lambda}^{a}=0$, which
means that every state is best matched relative to the previous one and
spatial points with the same $\vec{x}$ coordinates are equilocal. The
corresponding Hamiltonian, eq.(\ref{Hamiltonian c}), is
\begin{equation}
\tilde{H}_{0}[\psi,\psi^{\ast}]=\int dx\psi_{x}^{\ast}%
%TCIMACRO{\tsum \limits_{n}}%
%BeginExpansion
{\textstyle\sum\limits_{n}}
%EndExpansion
\left(  \frac{-\hbar^{2}}{2m_{n}}\delta^{ab}\partial_{na}\partial
_{nb}+V(\{|\vec{x}_{n}-\vec{x}_{m}|\})\right)  \psi_{x}~,
\label{Hamiltonian d}%
\end{equation}
and quantum states $\psi$ are restricted to the subspace constrained by
\begin{equation}
\tilde{P}_{a}[\psi,\psi^{\ast}]=0\quad\text{or}\quad\langle\psi|\hat{P}%
|\psi\rangle=0~. \label{BM constr b}%
\end{equation}

We conclude by noting that BM imposes expected value constraints. In the
standard approach to quantizing theories with constraints, questions arise as
to whether the constraints of the classical theory should, after quantization,
be imposed on the ontic microstates, on the operators, on the quantum states,
or on expectation values. The ED approach to BM provides a crisp answer: the
quantum constraints that express relationality are to be imposed on
expectation values.

\subsection{Best matching under rigid translations and rotations}

In this section we formulate a QM model that is relational with respect to the
ST3 transformations of both rigid translations and rotations, $\dot{\xi}%
_{n}^{a}=\varepsilon^{abc}\dot{\zeta}_{b}x_{nc}+\dot{\lambda}^{a}$. For
simplicity we shall assume that relationality with respect to translations has
already been imposed; the system is described in the center of mass frame,
$\dot{\lambda}^{a}=0$, and quantum states are constrained by
(\ref{BM constr b}). Then $\dot{\xi}_{n}^{a}=\varepsilon^{abc}\dot{\zeta}%
_{b}x_{nc}$ and the best-matching condition (\ref{BM a}) is written as
\begin{equation}
\frac{\partial}{\partial\dot{\zeta}_{a}}\tilde{H}_{\dot{\xi}}=0~. \label{BM d}%
\end{equation}
Substituting $\tilde{H}_{\dot{\xi}}$ from by (\ref{Hamiltonian c}), after some
straightforward algebra, we find%

\[
\frac{\partial\tilde{H}_{\dot{\xi}}}{\partial\dot{\zeta}_{a}}=-\int
dx\psi^{\ast}%
%TCIMACRO{\tsum \limits_{n}}%
%BeginExpansion
{\textstyle\sum\limits_{n}}
%EndExpansion
\varepsilon^{abc}x_{nb}(\frac{\hbar}{i}\partial_{nc})\psi-\int dx\psi^{\ast}%
%TCIMACRO{\tsum \limits_{n}}%
%BeginExpansion
{\textstyle\sum\limits_{n}}
%EndExpansion
m_{n}\varepsilon^{abd}\dot{\xi}_{nb}x_{nd}\psi~.
\]
The first integral on the right is recognized as the (expected) total angular
momentum,
\begin{equation}
\text{1st}=-\int dx\psi^{\ast}%
%TCIMACRO{\tsum \limits_{n}}%
%BeginExpansion
{\textstyle\sum\limits_{n}}
%EndExpansion
\varepsilon^{abc}x_{nb}(\frac{\hbar}{i}\partial_{nc})\psi=-\langle\psi|\hat
{L}^{a}|\psi\rangle=-\tilde{L}^{a}~. \label{int1}%
\end{equation}
Substituting $\dot{\xi}_{n}^{a}=\varepsilon^{abc}\dot{\zeta}_{b}x_{nc}$ into
the second integral,
\begin{equation}
\text{2nd}=-\int dx\psi^{\ast}%
%TCIMACRO{\tsum \limits_{n}}%
%BeginExpansion
{\textstyle\sum\limits_{n}}
%EndExpansion
m_{n}\varepsilon^{abc}\dot{\xi}_{nb}x_{n}^{c}\psi=\int dx\psi^{\ast}%
%TCIMACRO{\tsum \limits_{n}}%
%BeginExpansion
{\textstyle\sum\limits_{n}}
%EndExpansion
m_{n}\varepsilon^{bac}\varepsilon^{bde}\dot{\zeta}^{d}x_{n}^{e}x_{n}^{c}\psi
\end{equation}
and using the identity $\varepsilon^{bac}\varepsilon^{bde}=\delta^{ad}%
\delta^{ce}-\delta^{ae}\delta^{cd}$, we find%
\begin{equation}
\text{2nd}=\int dx\psi^{\ast}%
%TCIMACRO{\tsum \limits_{n}}%
%BeginExpansion
{\textstyle\sum\limits_{n}}
%EndExpansion
m_{n}(\delta^{ab}x_{n}^{c}x_{n}^{c}-x_{n}^{a}x_{n}^{b})\psi\dot{\zeta}%
_{b}=\langle\psi|\hat{I}^{ab}|\psi\rangle\dot{\zeta}_{b}=\tilde{I}^{ab}%
[\psi,\psi^{\ast}]\dot{\zeta}_{b}~, \label{int2}%
\end{equation}
where $\hat{I}^{ab}$ is recognized as the moment of inertia tensor and the
functional $\tilde{I}^{ab}[\psi,\psi^{\ast}]$ is its expected value. Combining
(\ref{int1}) and (\ref{int2}) we find
\begin{equation}
\frac{\partial\tilde{H}_{\dot{\xi}}}{\partial\dot{\zeta}_{a}}=\tilde{I}%
^{ab}\dot{\zeta}_{b}-\tilde{L}^{a}=0~. \label{BM constr c}%
\end{equation}
The interpretation is that if we are given a sequence of consecutive states,
the shift angular velocity $\dot{\zeta}^{a}$ that achieves minimal mismatch
with respect to rotations is given by the expected angular velocity of the
system,
\begin{equation}
\dot{\zeta}^{a}=(\tilde{I}^{-1})_{b}^{a}\tilde{L}^{b}\,. \label{ang vel}%
\end{equation}
Thus, in the center of mass frame ($\tilde{P}^{a}=0$) rotational relationality
is implemented through a BM constraint that restricts solutions to the
subspace of the full Hilbert space with a conserved expected total momentum
given by%
\begin{equation}
\tilde{L}^{a}=\tilde{I}^{ab}\dot{\zeta}_{b}~. \label{BM constr d}%
\end{equation}

In conclusion, to formulate a relational dynamics with evolution given by
(\ref{Hamiltonian c}) with $\dot{\xi}_{n}^{a}=\varepsilon^{abc}\dot{\zeta}%
_{b}x_{nc}+\dot{\lambda}^{a}$ chosen to achieve best matching, then we must
require that solutions be constrained to the subspace of the full Hilbert
space with total momentum and total angular momentum given by
\begin{equation}
\langle\psi|(\hat{P}-M\dot{\lambda}^{a})|\psi\rangle=0\quad\text{and}%
\quad\langle\psi|(\hat{L}^{a}-\hat{I}^{ab}\dot{\zeta}_{b})|\psi\rangle=0~.
\end{equation}
This completes our formulation of an ED model that is relational relative to
the rigid translations and rotations ST3.

So far the situation seems closely analogous to the case of pure translations
but in fact it is substantially different. While momentum conservation implies
that the center of mass velocity $\dot{\lambda}^{a}$ must be constant in time,
the conservation of angular momentum does not imply that the angular velocity
$\dot{\zeta}^{a}$ is constant because the system is not a rigid body and the
moment of inertia $\tilde{I}^{ab}=\langle\psi|\hat{I}^{ab}|\psi\rangle$ is not
constant in time. Obvious exceptions are those states for which $\tilde{L}%
^{a}=0$. Furthermore, if we want a dynamics that is relational with respect to
rotations then we ought to be able to transform to a rotating frame, but this
is not a symmetry, at least not obviously so: one cannot just transform to a
rotating frame and not expect the appearance of centrifugal forces. We meet
here the quantum analogue of Newton's bucket. The elegant way out of this
quandary has been known for a long time. The resolution is via Einstein's
equivalence principle: one can formulate a dynamics that is relational with
respect to rotations provided the centrifugal forces are interpreted as radial
gravitational forces. From this broadened perspective what is rotationally
relational is not the system of particles by themselves, but the composite
system of particles plus the gravitational field. This topic obviously
deserves a more detailed study but its pursuit would take us beyond the more
limited goals of this paper. Its further exploration will be taken up elsewhere.

\section{Formal developments}

\subsection{Action Principle}

In the ED framework the introduction of an action principle is not
fundamental; it is a merely convenient way to summarize the content of an
already established formalism. The usual procedure is reversed: given the
Schr\"{o}dinger equation (\ref{SchroEqn}) and the desired constraints, one
designs an action that reproduces them.

For simplicity, here we shall limit ourselves to the ST2 model and just
consider relationality with respect to rigid translations. The generalization
to include rotations is immediate. The starting point is to define the
differential
\begin{equation}
\delta\mathcal{A}=\int_{t_{1}}^{t_{2}}dt\,\int_{R}dx\,\left[  \delta\psi
^{\ast}\left(  i\hbar\partial_{t}\psi-\hat{H}_{\dot{\lambda}}\,\psi\right)
-\left(  i\hbar\partial_{t}\psi^{\ast}+\hat{H}_{\dot{\lambda}}^{\ast}%
\psi^{\ast}\right)  \delta\psi\right]
\end{equation}
where $\tilde{H}_{\dot{\lambda}}$ is given in (\ref{Hamiltonian c}) and the
variations $\delta\psi$ and $\delta\psi^{\ast}$ vanish at the boundary of the
region $R$. Then one integrates $\delta\mathcal{A}$ to get the action
\begin{equation}
\mathcal{A}=\int_{t_{1}}^{t_{2}}dt\,\left(  \int_{R}dx\,i\hbar\psi^{\ast
}\partial_{t}\psi-\tilde{H}_{\dot{\lambda}}(\,\psi,\psi^{\ast})\right)
\ .\label{action a}%
\end{equation}
Remarkably, it is not necessary to append additional terms to the action to
enforce constraints. Varying with respect to $\dot{\lambda}_{a}(t)$ already
gives the desired BM\ constraint, eq.(\ref{BM constr a}). In the action
formalism the mismatch correction $\dot{\lambda}_{a}$ plays the role of a
Lagrange multiplier.

\subsection{Parametrized Entropic Dynamics}

We have emphasized that the ED models developed here are already temporally
relational in the sense that there are neither external clocks nor an external
time. Our final goal is to make this temporal relationality explicit by
producing a model in which entropic time $t$ no longer plays the role of time
but, instead, becomes a dynamical variable on a par with $\psi_{x}$.

The technique is well known from the canonical formalism of general relativity
\cite{Kuchar 1973}\cite{Henneaux Teitelboim 1992}. It is called
\textquotedblleft parametrization\textquotedblright, and the resulting
theories are said to be \textquotedblleft generally
covariant\textquotedblright. The idea is to introduce an unphysical label
$x^{0}$ that retains one of the important roles of time, namely, that of
keeping track of the temporal ordering of states. Let
\begin{equation}
t=t(x^{0})\quad\text{and}\quad\partial_{0}t=\frac{\partial t}{\partial x^{0}}~
\end{equation}
be the velocity of entropic time $t$ with respect to the \textquotedblleft
time\textquotedblright\ label $x^{0}$. Then $\psi(x,t(x^{0}))$ and
$\dot{\lambda}(t(x^{0}))$ become functions of $x^{0}$ which we will just write
as $\psi(x,x^{0})$ and $\dot{\lambda}(x^{0})$. The action (\ref{action a})
becomes
\begin{equation}
\mathcal{A}=\int_{t_{1}}^{t_{2}}dx^{0}\,\left(  \int_{R}dx\,i\hbar\psi^{\ast
}\partial_{0}\psi-\partial_{0}t\,\tilde{H}_{\dot{\lambda}}(\,\psi,\psi^{\ast
})\right)  \ .
\end{equation}
Note that $\mathcal{A}$ is linear in both the $\psi$ velocity and the $t$
velocity. This suggests that we can reinterpret $t$ as one of the canonical
coordinates provided we simultaneously reinterpret $-\tilde{H}_{\dot{\lambda}%
}(\,\psi,\psi^{\ast})=\pi_{0}$ as its conjugate momentum. The new action,
\begin{equation}
\mathcal{A}=\int_{t_{1}}^{t_{2}}dx^{0}\,\left(  \pi_{0}\partial_{0}t+\int%
_{R}dx\,i\hbar\psi^{\ast}\partial_{0}\psi\right)  \ ,
\end{equation}
is in standard $%
%TCIMACRO{\tint }%
%BeginExpansion
{\textstyle\int}
%EndExpansion
p\dot{q}$ form. The canonical variables $\{t,\pi_{0},\psi,i\hbar\psi^{\ast}\}$
cannot, however, be varied independently because they are subject to the
so-called super-Hamiltonian constraint,
\begin{equation}
\mathcal{\tilde{H}}\overset{\text{def}}{=}\pi_{0}+\tilde{H}_{\dot{\lambda}%
}(\,\psi,\psi^{\ast})=0~,\label{super H}%
\end{equation}
which must, therefore, be appended to the action with its corresponding
Lagrange multiplier,
\begin{equation}
\mathcal{A}=\int_{t_{1}}^{t_{2}}dx^{0}\,\left(  \pi_{0}\partial_{0}t+\int%
_{R}dx\,i\hbar\psi^{\ast}\partial_{0}\psi-\beta\mathcal{\tilde{H}}\right)
\ .\label{action b}%
\end{equation}
Now we can vary all $\{t,\pi_{0},\psi,i\hbar\psi^{\ast}\}$ independently and
we can proceed to obtain equations of motion where $\beta\mathcal{\tilde{H}}$
plays the role of a Hamiltonian. In particular,
\begin{equation}
\partial_{0}t=\frac{\partial}{\partial\pi_{0}}\beta\mathcal{\tilde{H}}=\beta~,
\end{equation}
allows us to interpret $\beta$ as the rate of change of entropic time $t$ with
respect to the label $x^{0}$, a quantity usually called the lapse function
\cite{ADM 1962}.

Constraints often generate invariances and the super-Hamiltonian
$\mathcal{\tilde{H}}$ is no different. An important consequence of including
$t$ among the dynamical variables is that the action (\ref{action b}) turns
out to be invariant under the relabeling
\begin{equation}
x^{0}\rightarrow\bar{x}^{0}=\bar{x}^{0}(x^{0})~. \label{GenCov}%
\end{equation}
The proof is immediate; just note that
\begin{equation}
dx^{0}\,\beta=dx^{0}\partial_{0}t=d\bar{x}^{0}\partial_{\bar{0}}t=d\bar{x}%
^{0}\,\bar{\beta}~.
\end{equation}
The invariance with respect to the continuous gauge group (\ref{GenCov}) is
referred to as general covariance.

To highlight the potential significance of the results above we conclude with
the observation that in the canonical quantization of gravity the
super-Hamiltonian of the classical theory, $\mathcal{H}_{\text{class}}$, is
constrained to vanish weakly (\emph{i.e.} vanish on the manifold of solutions
of the equations of motion) and this gives rise to the famous problem of time
(see \emph{e.g.}, \cite{Kiefer 2007}). The problem is that upon quantization
the quantum version of the constraint takes the form $\hat{H}\Psi=0$, the
quantum Hamiltonian annihilates the physical quantum states. The
Wheeler-DeWitt equation --- the analogue of the Schr\"{o}dinger equation ---
then implies that physical states cannot evolve.

Buckets of effort and ingenuity have been poured on this issue (see
\emph{e.g.}, \cite{Kuchar 2011}) and we shall not muddy it further. Our sole
concern here is to point out that ED evades all these problems because it
constructs the quantum theory directly without first formulating a classical
theory that must then be massaged according to quantization rules that to this
day elicit controversy.

\section{Conclusions}

The ED approach to QM has been extended to formulate non-relativistic quantum
models that are spatially relational with respect to rigid translations and
rotations. Although only partially relational --- absolute structures of
simultaneity and Euclidean geometry are retained --- our models still provide
a useful testing ground for ideas that will prove useful in the context of
more realistic relativistic theories.

The fact that within ED the positions of particles are meant to be ontic with
definite values at all times, just as in classical mechanics, has allowed us
to adapt and adopt some intuitions from Barbour and Bertotti's classical
framework. Nevertheless, the two frameworks are very different. Their
classical measure of mismatch compares ontic particle configurations and is
based on Jacobi's action. In contrast, our quantum measure of mismatch
compares epistemic quantum states and is adapted to the metric and symplectic
structures of the epistemic phase space.

The case of a single particle serves to illustrate the difference. While it
makes no sense to consider the relational classical motion of a single
particle, in the quantum case a relational dynamics makes sense even for a
single particle because what are being best-matched are infinite dimensional
wave functions.

We have shown that even in its previous formulations ED was already temporally
relational. Here we made this feature explicit and, as an example of a
development that might prove valuable in the context of quantum gravity, the
non-relativistic quantum model was rewritten in generally covariant form. This
toy model shows that the ED approach evades the analogue of what in quantum
gravity has been called the problem of time.

Finally, we note that the relational ED framework developed here can be
applied beyond the example of particles. It is expected to apply to any model
with redundancy in description, and this potentially includes all fundamental
theories such as electromagnetism, Yang-Mills theories and, possibly, gravity.

\subsubsection*{Acknowledgements}

AC would like to express gratitude to Selman Ipek for many discussions in the
early stages of this project. HS wants to thank Abdul Afzal and Hamza Waseem
for several insightful conversations about the foundations of quantum
mechanics. Both AC and HS would like to acknowledge an insightful discussion
with Julian Barbour.

\end{document}